\documentclass[a4paper,onecolumn]{IEEEtran}

\usepackage{amsmath,amssymb,amsfonts}
\usepackage{mathrsfs}
\usepackage{graphicx}
\usepackage{tabularx}
\usepackage{subcaption}
\usepackage{textcomp}
\usepackage{smartdiagram}
\usepackage{adjustbox}
\usepackage{pgfplots}
\usepackage{float}
\usepackage{xcolor}
\usepackage{bm}
\usepackage[boxed,commentsnumbered, linesnumbered,ruled,vlined]{algorithm2e}
\usepackage[sorting=none]{biblatex}
\def\BibTeX{{\rm B\kern-.05em{\sc i\kern-.025em b}\kern-.08em
    T\kern-.1667em\lower.7ex\hbox{E}\kern-.125emX}}
\usepackage{epstopdf}
\usepackage{siunitx}
\usepackage{tikz}
\usepackage{circuitikz}
\usepackage{dirtytalk}
\usepackage{svg}
\usepackage{listings}
\usetikzlibrary{shapes.geometric, arrows, positioning}
\usepackage{titlesec}
\titlespacing{\subsubsection}{0pt}{10pt}{5pt}

\usepackage[normalem]{ulem}

\usepackage[left=2.2cm, right=2.2cm, top=3.5cm, bottom=2.5cm]{geometry}

\usepackage[skip=5pt plus1pt, indent=10pt]{parskip}

\usepackage{fancyhdr}
\graphicspath{{./figures/}}

\tikzstyle{startstop} = [rectangle, rounded corners, minimum width=3cm, minimum height=1cm,text centered, draw=black, fill=red!30]
\tikzstyle{io} = [trapezium, trapezium left angle=70, trapezium right angle=110, minimum width=3cm, minimum height=1cm, text centered, draw=black, fill=blue!30]
\tikzstyle{process} = [rectangle, minimum width=3cm, minimum height=1cm, text centered, text width=4cm, draw=black, fill=orange!30]
\tikzstyle{decision} = [diamond, minimum width=3cm, minimum height=1cm, text centered, draw=black, fill=green!30]
\tikzstyle{arrow} = [thick,->,>=stealth]

\definecolor{ForestGreen}{RGB}{34,150,34}

\begin{document}

\title{Efficient Verification of a RADAR SoC Using Formal and Simulation-Based Methods}

\author{
\IEEEauthorblockA{\vspace{4mm}Aman Kumar,
Infineon Technologies,
Dresden, Germany
(\textit{aman.kumar@infineon.com})}

\IEEEauthorblockA{Mark Litterick,
Verilab GmbH,
Munich, Germany
(\textit{mark.litterick@verilab.com})}

\IEEEauthorblockA{Samuele Candido,
Infineon Technologies,
Dresden, Germany
(\textit{samuele.candido@infineon.com})}}

\maketitle

\thispagestyle{fancy}


\begin{abstract}
\textbf{\emph{Abstract}\!
\textemdash As the demand for Internet of Things (IoT) and Human-to-Machine Interaction (HMI) increases, modern System-on-Chips (SoCs) offering such solutions are becoming increasingly complex. This intricate design poses significant challenges for verification, particularly when time-to-market is a crucial factor for consumer electronics products. This paper presents a case study based on our work to verify a complex Radio Detection And Ranging (RADAR) based SoC that performs on-chip sensing of human motion with millimetre accuracy \cite{saverio}. We leverage both formal and simulation-based methods to complement each other and achieve verification sign-off with high confidence \cite{formal_sim}. While employing a requirements-driven flow approach \cite{rddf}, we demonstrate the use of different verification methods to cater to multiple requirements and highlight our know-how from the project. Additionally, we used Machine Learning (ML) based methods, specifically the Xcelium ML tool from Cadence, to improve verification throughput \cite{deepak_ml}.}
\end{abstract}


\section{Introduction}
\label{sec:intro}

Verification has become the bottleneck in product development cycles, as it takes more than \SI{60}{\percent} of the overall project time \cite{VerStudy}. Complex designs such as RADAR-based SoC contribute even more to the challenges in verification on top of existing ones. Our RADAR SoC is a tiny device that has analog and digital domains that collect data from different sensors to act as a human-machine interface.

\begin{figure}[h!]
\centering
\begin{tikzpicture}
\begin{axis}[
    ylabel={Percentage of Design Projects},
    xmin=0, xmax=100,
    ymin=0, ymax=40,
    xtick={0,10,20,30,40,50,60,70,80,90,100},
    xticklabels={,More than \SI{10}{\percent} early,\SI{10}{\percent},On-schedule,\SI{10}{\percent} behind schedule,\SI{20}{\percent},\SI{30}{\percent},\SI{40}{\percent},\SI{50}{\percent},$>$\SI{50}{\percent} behind schedule},
    x tick label style={font=\footnotesize,rotate=45, anchor=east},
    ytick={0,10,20,30,40},
    legend pos=north east,
    ymajorgrids=true,
    grid style=dashed,
]

\addplot[
    color=cyan,
    mark=triangle,
    line width=0.25mm,
    ]
    coordinates {
    (10,1)(20,1)(30,32)(40,35)(50,14)(60,8)(70,4)(80,2)(90,5)
    };
    \addlegendentry{2010}

\addplot[
    color=darkgray,
    mark=square,
    line width=0.25mm,
    ]
    coordinates {
    (10,2)(20,4)(30,32)(40,28)(50,14)(60,8)(70,4)(80,4)(90,8)
    };
    \addlegendentry{2014}

\addplot[
    color=green,
    mark=*,
    line width=0.25mm,
    ]
    coordinates {
    (10,1.5)(20,2)(30,27)(40,29)(50,18)(60,9)(70,4)(80,4)(90,5)
    };
    \addlegendentry{2018}

\addplot[
    color=orange,
    mark=+,
    line width=0.25mm,
    ]
    coordinates {
    (10,1.4)(20,1.8)(30,28)(40,26)(50,15)(60,15)(70,4)(80,4.5)(90,6)
    };
    \addlegendentry{2022}

\end{axis}
\end{tikzpicture}
\caption{Actual ASIC design completion compared to project's original schedule \cite{VerStudy}}
\label{asic_completion_time}
\end{figure}
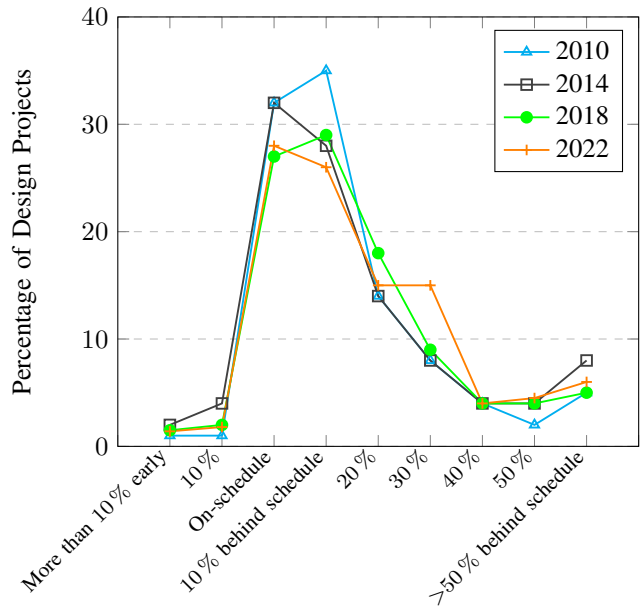

The study in Fig.~\ref{asic_completion_time} \cite{VerStudy} shows that around \SI{66}{\percent} of Application Specific Integrated Circuit (ASIC) projects are behind schedule and \SI{27}{\percent} of projects are behind the schedule by \SI{30}{\percent} or more. One of the significant contributors to the schedule not being met is functional verification, which could be expensive for a consumer electronics product. To overcome the verification challenges for a product that requires a faster Time to Market (TTM), we used different verification techniques to verify different functionalities. The project uses a requirements-based approach to collect all design requirements and create a corresponding verification plan (vPlan) to cover the requirements. We use a SystemVerilog-UVM based testbench to verify the SoC top-level with individual UVM Verification Components (UVCs) for different subsystems. Since the usage of formal property verification and automated formal checks is increasing which brings real benefit to the project \cite{VerStudy}, we use formal verification to verify complex features and blocks that are formal-friendly \cite{foster_guidelines} \cite{fvbook}. We use Formal Property Verification (FPV) to verify design behaviours, Connectivity (CONN) verification to check end-to-end conditional connections, Control/Status Register (CSR) verification to verify the register blocks, Assertion-Based Verification IPs (ABVIPs) to verify standard protocols such as AHB and Unreachability (UNR) analysis to improve faster code coverage closure. We also perform power aware simulations to verify different power domains and Gate-Level Simulations (GLS) to confirm the correct design behaviour on the netlist level. On top of all the verification methods, we use the Xcelium ML tool from Cadence to achieve optimized regression with efficient functional and code coverage.

\section{Design Overview}

\begin{figure}[h!]
\centering
  \includegraphics [width=0.5\textwidth] {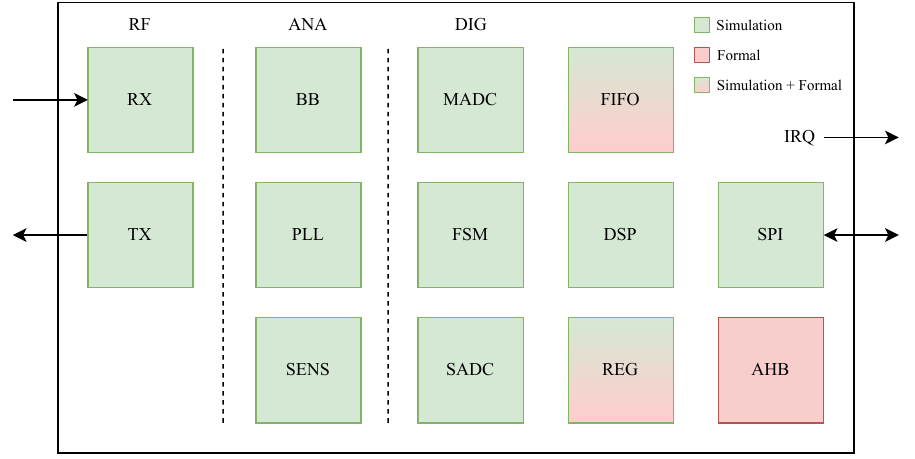}
\caption{RADAR SoC block diagram}
\label{soc_block_dig}
\end{figure}

A simplified block diagram of the RADAR SoC is represented in Fig.~\ref{soc_block_dig}. The SoC provides all the functionality required for a single-chip RADAR sensor device. It consists of Radio Frequency (RF) Transmit (TX) and Receive (RX) blocks which provide an air-side interface via the in-package antennas.  Several key Analog (ANA) blocks provide the Phase-Locked-Loop (PLL) for the TX frequency ramps and base-band processing of the RX signals. In addition, there are embedded sensors for power and temperature measurement (among others). The Digital (DIG) part of the device includes Multichannel Analog-to-Digital Converter (MADC) and Sensor ADC (SADC) blocks, microcoded control engine (Finite-State Machine (FSM)), Digital Signal Processing (DSP) capability, memory, First In First Out (FIFO), register, and power management blocks connected via an AHB bus. The main interface to the host is via a Serial Peripheral Interface (SPI) and discrete interrupt and trigger signals.

\section{Design Verification}
\label{sec:dv}

These RADAR SoC devices are primarily targeting portable consumer applications, resulting in the inevitable pressure on project timelines. From an architectural perspective, the main challenges facing the verification process include the high degree of analog content in the device, demanding low-power performance criteria, complicated and flexible control operation, and complex mathematical signal processing. To meet our design verification goals, a variety of verification methods were used for different blocks (simulation, formal or a mixture), as shown in Fig.~\ref{soc_block_dig}. The following sub-sections provide an overview of the techniques used.

\subsection{Simulation Based Verification}

\begin{figure}[h!]
\centering
  \includegraphics [width=0.8\textwidth] {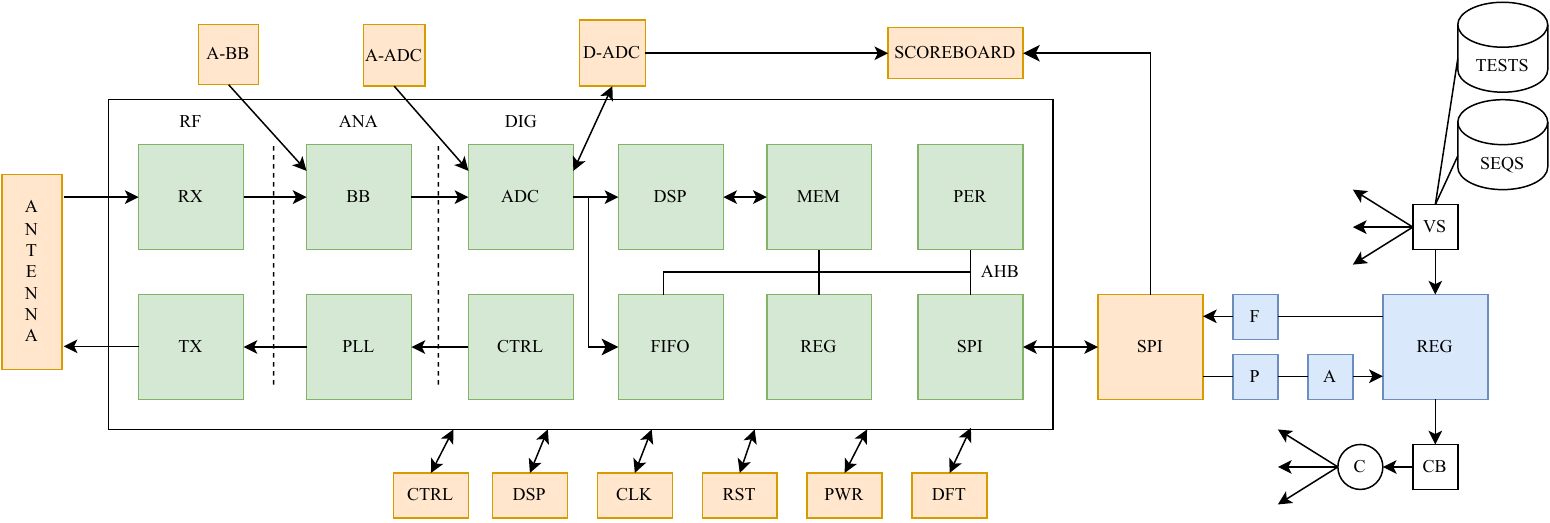}
\caption{UVM testbench block diagram}
\label{tb_block_dig}
\end{figure}

The simulation testbench is written in SystemVerilog using UVM. The same testbench and test suite are used to support digital, power-aware, gate-level and (indirectly) analog/mixed-signal simulations. A simplified block diagram of the UVM testbench is shown in Fig.~\ref{tb_block_dig}. The testbench consists of multiple UVCs, each with responsibility for active stimulus on external interfaces, internal analog or digital streaming capabilities, or internal passive monitoring. The environment includes an automatically generated UVM Register (REG) model, which is integrated to the SPI UVC using Frontdoor (F), Adapter (A) and Predictor (P) components. The environment uses Callbacks (CB) to maintain the configuration status in each UVC in response to register operations on the Design Under Verification (DUV) \cite{reg_model}. The DUV in the UVM environment uses real-number models for the corresponding analog blocks (TX, RX, BB, part PLL, and part ADC). The test stimulus is coded in sequences that run on the top-level Virtual Sequencer (VS) and make use of a constrained-random sequence library and low-level sequences belonging to each UVC. The main data path is validated by means of a scoreboard UVC which is connected to ADC and SPI components, but there are also signal protocol checks in each UVC interface as well as self-checking mechanisms in all tests. Functional coverage is distributed throughout the environment.

\subsubsection{Digital Simulation}

The digital UVC agents for external connections to Clock (CLK), Reset (RST), Power (PWR) and SPI all provide a set of sequences for stimulus generation. Analog model stimulus for RX, BB and ADC components is provided by configuration sequences and real-number streaming mechanisms (which force real number values and patterns onto internal analog nets) using the streaming techniques described in \cite{data_streaming}. Most of the digital tests use the corresponding ADC inputs as the main path stimulus (MADC or SADC), but a few device features require BB or antenna stimulus as the source.

The DSP block requires some advanced verification techniques in order to validate the mathematical algorithms embedded in the hardware. These techniques include combining design-for-verification modes with a known-answer-test strategy and using pre-defined data scenarios that are generated externally by MatLab to predict results for specific configurations. These tests are supplemented by constrained random exploration of the configuration space at a later stage in the project once additional mathematical models are available. In these tests, the raw data source values from the MatLab scenarios are streamed directly onto the ADC digital outputs using a technique that is also described in \cite{data_streaming}.

In order to manage the linear development of the design while the verification environment evolved and make debugging more effective, the sequence hierarchy is carefully architected to enable isolation and combination of features using the techniques described in \cite{sequence_pro}. Specifically, we isolate individual features into independent exhaustive tests, and separately combine relevant features in additional compound tests, ultimately providing a few tests that combine all features with limited random exploration of the configurations space. With this goal in mind, the main UVM testcases are divided into the following groups:

\begin{itemize}
    \item Basic tests: verify basic features and paths with limited randomization of the stimulus
    \item Feature tests: verify each feature with comprehensive constrained-random stimulus
    \item Stress tests: verify stress conditions that could reasonably be expected in the application
    \item Usecases tests: verify real application scenarios defined by concept, application or customer
\end{itemize}

Note that stress tests are not intended to address all Garbage-In/Garbage-Out (GIGO) scenarios such as incorrect configuration, but rather illegal conditions that could reasonably be expected to occur in real life (e.g. length errors on SPI, data path overflow and underrun, trigger mismanagement and interrupt handling errors, etc.). In addition, we also have dedicated tests for Design-for-Test (DFT) modes and provide support for re-simulation of test equipment scenarios to assist software development, post-silicon validation, and test-engineering teams.

\subsubsection{Power Aware Simulation}

Standard RTL simulations are used for digital verification. These simulations are not power-aware; however, power control signal protocol checks are embedded within relevant UVC agents (e.g. clock, power, control FSM) and are active during all simulations. Full power-aware simulations using the Unified Power Format (UPF) standard are supported using a run-time command (and dedicated regression). These simulations run the same tests on the same environment but enable UPF operation and therefore power-awareness during simulation.

Low power consumption is a key element that allows the RADAR to be adopted in consumer and IoT applications \cite{saverio}. Power gating is one of the techniques adopted in this architecture, in order to reduce power consumption. Partitioning the design in multiple power domains allows to switch off parts of the chip that are not needed. The power management task is taken over by a power domain controller, which drives the power switches and enables and disables isolation for a certain power domain.

In a traditional RTL simulation approach, important aspects of the power strategy cannot be verified. Power ground nets, power switches, and isolation cells are among the power aspects not considered in traditional RTL simulations. In order to identify power management issues and potential bugs in the early stage of the project, power-aware verification becomes highly relevant. The most common approach is based on the IEEE Std 1801 UPF, which allows one to describe, model, and specify the power intent of the design. Multiple UPF files are used in the RADAR SoC.  For each IP in the design, a UPF file is provided that describes its power intention. In addition to that, a main UPF defines the power domains, as well as the power switches; creates and connects the power supplies and loads the UPF files of IP modules.

First, a testbench environment was created for the UPF simulation of RTL with Cadence Xcelium. After that, a simple simulation was run to verify the basic RADAR functionality. Between this phase and the passing of the first test, we faced a few challenges. First of all: if the main UPF file was written for synthesis tools, there might be differences in the paths to design instances or signals. This is particularly true for module instances created with generate statements. This kind of issue usually causes errors during compilation. A more serious issue is the isolation of SystemVerilog interfaces, which are intensively used in RADAR design as ports on the power domain boundaries. The support for SystemVerilog interface isolation is limited to the Xcelium version adopted in the project. The isolation of interface elements of type {\fontfamily{qcr}\selectfont enum} is, for example, not supported. In order to overcome this issue, which led to failing simulations, workarounds had to be used. In this case, the UVM routine {\fontfamily{qcr}\selectfont uvm\_hdl\_deposit} was used to mimic the isolation behaviour. With such workarounds great care is required since they might hide functional issues. After solving the mentioned issues and other secondary problems, all related to the setup and the tools, the RTL UPF simulation showed its benefits. Missing isolation and, therefore, propagation of X's and wrong switch-off sequence are two prominent examples. It is worth mentioning that this kind of bug could have probably been found in GLS. However, GLS is normally performed in a phase close to tape-out and the time and effort for debugging would have been much higher.

\subsubsection{Gate-Level Simulation}

We have a requirement to perform GLS not just to validate the post-synthesis netlist and cross-check the static-timing-analysis constraints, but also to enable us to do more accurate power analysis on the cell-based netlist after clock-tree synthesis. This power analysis is required for low-power and high-power evaluation and must be repeated with various timing corners (e.g. nominal, slow, and fast). It is not the goal of GLS to prove all low-level protocol and functional operations are still intact (this is supported by equivalence checking), but we do need to know if tests are working correctly in order to be sure that the test ran correctly.

In order to minimize the burden of GLS on the tight timescales, we applied a strategy to minimize testbench effort and reduce maintenance between release candidates. Specifically, since all our tests are self-checking (they all call result and status sequences to validate important aspects of the results), we implement a mechanism to disconnect all internal passive UVC interfaces completely during GLS. All external UVCs (such as SPI, CLK, RST, and PWR) remain connected, as too does the ADC interface (since it is the primary input for most tests) and the primary data-path scoreboard and a few critical clock and power checks. This results in a need to maintain far fewer internal connects to the netlist, which in turn provides the back-end team with more flexibility in terms of flattening and processing the netlist (i.e., we have far fewer nets that we need to find in the gate-level netlist). 

This technique allows us to run almost all tests out of the box with very little maintenance overhead. The same suite of tests is run on exactly the same UVM environment, where the only difference is that most UVCs are disconnected, and some key remaining nets need alternative paths to be defined for the netlist. Note that in general the HDL-paths for the register model connections are no longer valid in GLS after netlist optimization, so a few tests that rely on these paths to do backdoor operations are not executed in GLS regressions.

\subsubsection{Analog/Mixed-Signal Simulation}

As mentioned above, the digital simulations are run with real-number models for the analog macros instantiated within the DUV. These models are provided by the analog design team. In order to run high-level analog and Analog/Mixed-Signal (AMS) simulations on the real analog netlist, we simulate dedicated AMS usecase tests on the digital UVM simulation environment and export the corresponding Value Change Dump (VCD) file to our analog team. They use these VCD files as direct digital stimuli for the full DUV with analog netlist in an appropriate analog simulator setup. In this respect, the UVM environment supports AMS indirectly.

\subsection{Formal Verification}

Formal verification uses technologies that mathematically analyze the space of possible behaviours of a design, rather than computing results for particular values \cite{fvbook}. It is an exhaustive verification technique that uses mathematical proof methods to verify whether the design implementation matches design specifications \cite{aman_dvcon_ecc}.

\tikzset{every picture/.style={line width=1pt}} 
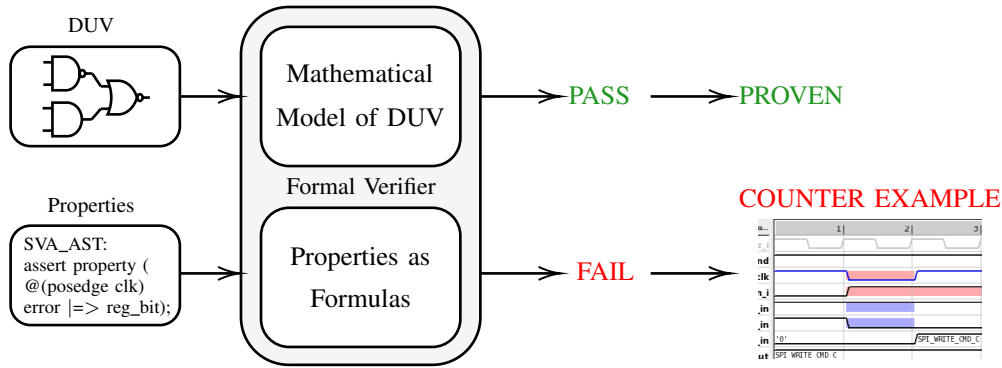
\begin{figure}[h]
\centering
\begin{tikzpicture}[x=0.75pt,y=0.75pt,yscale=-1,xscale=1]

\draw [fill={rgb, 255:red, 155; green, 155; blue, 155 }  ,fill opacity=0.1 ]  (271,133.9) .. controls (271,120.7) and (281.7,110) .. (294.9,110) -- (366.6,110) .. controls (379.8,110) and (390.5,120.7) .. (390.5,133.9) -- (390.5,266.1) .. controls (390.5,279.3) and (379.8,290) .. (366.6,290) -- (294.9,290) .. controls (281.7,290) and (271,279.3) .. (271,266.1) -- cycle ;
\draw [fill={rgb, 255:red, 255; green, 255; blue, 255 }  ,fill opacity=1 ] (281,134) .. controls (281,126.27) and (287.27,120) .. (295,120) -- (366.5,120) .. controls (374.23,120) and (380.5,126.27) .. (380.5,134) -- (380.5,176) .. controls (380.5,183.73) and (374.23,190) .. (366.5,190) -- (295,190) .. controls (287.27,190) and (281,183.73) .. (281,176) -- cycle ;
\draw [fill={rgb, 255:red, 255; green, 255; blue, 255 }  ,fill opacity=1 ] (281,225) .. controls (281,217.27) and (287.27,211) .. (295,211) -- (366.5,211) .. controls (374.23,211) and (380.5,217.27) .. (380.5,225) -- (380.5,267) .. controls (380.5,274.73) and (374.23,281) .. (366.5,281) -- (295,281) .. controls (287.27,281) and (281,274.73) .. (281,267) -- cycle ;
\draw   (156,140) .. controls (156,134.48) and (160.48,130) .. (166,130) -- (230.5,130) .. controls (236.02,130) and (240.5,134.48) .. (240.5,140) -- (240.5,170) .. controls (240.5,175.52) and (236.02,180) .. (230.5,180) -- (166,180) .. controls (160.48,180) and (156,175.52) .. (156,170) -- cycle ;
\draw   (156,230) .. controls (156,224.48) and (160.48,220) .. (166,220) -- (230.5,220) .. controls (236.02,220) and (240.5,224.48) .. (240.5,230) -- (240.5,260) .. controls (240.5,265.52) and (236.02,270) .. (230.5,270) -- (166,270) .. controls (160.48,270) and (156,265.52) .. (156,260) -- cycle ;
\draw    (240.5,155) -- (268.5,155) ;
\draw [shift={(270.5,155)}, rotate = 180] [color={rgb, 255:red, 0; green, 0; blue, 0 }  ][line width=0.75]    (10.93,-3.29) .. controls (6.95,-1.4) and (3.31,-0.3) .. (0,0) .. controls (3.31,0.3) and (6.95,1.4) .. (10.93,3.29)   ;
\draw    (240.5,244) -- (268.5,244) ;
\draw [shift={(270.5,244)}, rotate = 180] [color={rgb, 255:red, 0; green, 0; blue, 0 }  ][line width=0.75]    (10.93,-3.29) .. controls (6.95,-1.4) and (3.31,-0.3) .. (0,0) .. controls (3.31,0.3) and (6.95,1.4) .. (10.93,3.29)   ;
\draw   (176.89,135) -- (183.69,135) .. controls (187.43,135) and (190.48,138.79) .. (190.48,143.45) .. controls (190.48,148.11) and (187.43,151.9) .. (183.69,151.9) -- (176.89,151.9) -- (176.89,135) -- cycle (172.36,137.82) -- (176.89,137.82) (172.36,149.08) -- (176.89,149.08) (193.2,143.45) -- (196.82,143.45) (190.48,143.45) .. controls (190.48,142.52) and (191.09,141.76) .. (191.84,141.76) .. controls (192.59,141.76) and (193.2,142.52) .. (193.2,143.45) .. controls (193.2,144.38) and (192.59,145.14) .. (191.84,145.14) .. controls (191.09,145.14) and (190.48,144.38) .. (190.48,143.45) -- cycle ;
\draw   (204.44,147.11) -- (210.1,147.11) .. controls (214.05,147.26) and (217.58,150.56) .. (219.16,155.56) .. controls (217.58,160.57) and (214.05,163.86) .. (210.1,164.01) -- (204.44,164.01) .. controls (206.87,158.78) and (206.87,152.34) .. (204.44,147.11) -- cycle (201.04,149.93) -- (205.57,149.93) (201.04,161.2) -- (205.57,161.2) (221.88,155.56) -- (225.5,155.56) (219.16,155.56) .. controls (219.16,154.63) and (219.77,153.87) .. (220.52,153.87) .. controls (221.27,153.87) and (221.88,154.63) .. (221.88,155.56) .. controls (221.88,156.5) and (221.27,157.25) .. (220.52,157.25) .. controls (219.77,157.25) and (219.16,156.5) .. (219.16,155.56) -- cycle ;
\draw   (176.89,158.1) -- (184.23,158.1) .. controls (188.28,158.1) and (191.57,161.89) .. (191.57,166.55) .. controls (191.57,171.21) and (188.28,175) .. (184.23,175) -- (176.89,175) -- (176.89,158.1) -- cycle (172,160.92) -- (176.89,160.92) (172,172.18) -- (176.89,172.18) (191.57,166.55) -- (196.46,166.55) ;
\draw    (201.04,161.2) -- (196.46,166.55) ;
\draw    (201.04,149.93) -- (196.82,143.45) ;

\draw    (390.5,155) -- (428.5,155) ;
\draw [shift={(430.5,155)}, rotate = 180] [color={rgb, 255:red, 0; green, 0; blue, 0 }  ][line width=0.75]    (10.93,-3.29) .. controls (6.95,-1.4) and (3.31,-0.3) .. (0,0) .. controls (3.31,0.3) and (6.95,1.4) .. (10.93,3.29)   ;
\draw    (390.5,244) -- (428.5,244) ;
\draw [shift={(430.5,244)}, rotate = 180] [color={rgb, 255:red, 0; green, 0; blue, 0 }  ][line width=0.75]    (10.93,-3.29) .. controls (6.95,-1.4) and (3.31,-0.3) .. (0,0) .. controls (3.31,0.3) and (6.95,1.4) .. (10.93,3.29)   ;
\draw    (475.5,155) -- (513.5,155) ;
\draw [shift={(515.5,155)}, rotate = 180] [color={rgb, 255:red, 0; green, 0; blue, 0 }  ][line width=0.75]    (10.93,-3.29) .. controls (6.95,-1.4) and (3.31,-0.3) .. (0,0) .. controls (3.31,0.3) and (6.95,1.4) .. (10.93,3.29)   ;
\draw    (475.5,244) -- (513.5,244) ;
\draw [shift={(515.5,244)}, rotate = 180] [color={rgb, 255:red, 0; green, 0; blue, 0 }  ][line width=0.75]    (10.93,-3.29) .. controls (6.95,-1.4) and (3.31,-0.3) .. (0,0) .. controls (3.31,0.3) and (6.95,1.4) .. (10.93,3.29)   ;
\draw (585,252) node  {\includegraphics[width=83.63pt,height=52.5pt]{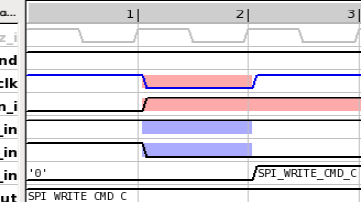}};

\draw (292,194) node [anchor=north west][inner sep=0.75pt]  [font=\small] [align=left] {Formal Verifier};
\draw (292,137) node [anchor=north west][inner sep=0.75pt]   [align=left] {Mathematical};
\draw (287,159) node [anchor=north west][inner sep=0.75pt]   [align=left] {Model of DUV};
\draw (294,229) node [anchor=north west][inner sep=0.75pt]   [align=left] {Properties as};
\draw (304,250) node [anchor=north west][inner sep=0.75pt]   [align=left] {Formulas};
\draw (183,114) node [anchor=north west][inner sep=0.75pt]  [font=\footnotesize] [align=left] {DUV};
\draw (161,224) node [anchor=north west][inner sep=0.75pt]  [font=\scriptsize] [align=left] {SVA\_AST:\\assert property (\\@(posedge clk)\\error $|$=$>$ reg\_bit);};
\draw (173,204) node [anchor=north west][inner sep=0.75pt]  [font=\footnotesize] [align=left] {Properties};
\draw (433,148) node [anchor=north west][inner sep=0.75pt]   [align=left] {\textcolor{ForestGreen}{PASS}};
\draw (437,237) node [anchor=north west][inner sep=0.75pt]   [align=left] {\textcolor{red}{FAIL}};
\draw (518,148) node [anchor=north west][inner sep=0.75pt]   [align=left] {\textcolor{ForestGreen}{PROVEN}};
\draw (518,200) node [anchor=north west][inner sep=0.75pt]   [align=left] {\textcolor{red}{COUNTER EXAMPLE}};

\end{tikzpicture}
\caption{Formal verifier \cite{aman_dvcon_ecc}}
\label{formal_verifier}
\end{figure}

Fig.~\ref{formal_verifier} shows the working of a formal verifier. There are two inputs to the formal verifier tool. On the one hand, the Design Under Verification (DUV) is fed into the tool which is converted into a mathematical model. On the other hand, properties, written in SystemVerilog Assertions (SVA) that capture the intent of the design are fed into the tool. The tool then converts these properties into mathematical formulas. In the next step, the tool tries to prove these mathematical formulas on the mathematical model of the DUV. If the properties do not hold, it is said to have failed, and a Counter Example (CEX) is generated by the tool to further debug. In general, the absence of a CEX is nothing but a pass or proven result \cite{aman_dvcon_configvermet}. To meet different verification requirements, we used different formal verification techniques and apps from Cadence Jasper. The following sub-sections provide an overview of the techniques used.

\subsubsection{Control/Status Register Verification}

The CSR verification app from Cadence Jasper offers a pragmatic solution for verifying the registers in the DUV. It is especially beneficial at the early stage of the project when the UVM testbench is not fully functional yet we need to verify the correct behaviour of the registers. The register description is prepared by the concept engineer and is used to generate the Comma Separated Values (CSV) file, which is an input to the formal tool. This generation is based on a metamodel-based automation framework explained in \cite{metamodel}. Since this approach is fully automated and a push-button solution, it also helps design engineers verify their RTL for every register change before releasing the next version of the RTL.

A major benefit of using formal based approach for register verification is the ease of debugging. The CEXs in case of a failure are usually within 10 clock cycles which helps to identify the root cause easily. On the other hand, a failure in simulation based setup may happen after several clock cycles, making the debugging of the root cause difficult. However, there are also limitations associated with this approach. Usually, the properties generated by the tool are encrypted and cannot be accessed. On the other hand, custom register schemes that are specific to the project cannot be verified with this approach as the properties are only generated for standard register types.

\subsubsection{Connectivity Verification}

The CONN app from Cadence Jasper offers formal based connectivity verification for straight and conditional connections. Traditional simulation based connectivity checking does not exercise a brute force approach; however, formal verification gives complete coverage due to its exhaustive nature. In our design, we have a lot of digital multiplexer (DMUX) connections that need to be verified. We have several analog blocks that connect to the digital top. We verify end-to-end connectivity for such connections. We also have several overwrite bits in the test registers that are used during debug mode. These overwrite bits are also a good candidate for such a connectivity check that we performed for the project.

Similar to the DMUX, the DUV also has many Analog Multiplexer (AMUX) which we wanted to verify with the formal approach. However, since most modern formal tools only support synthesizable designs and not analog/AMS models, it was not possible to verify the AMUX using this approach. We had to use simulation based approach to verify the design in this case. Although there are several techniques to also verify connectivity in analog/AMS designs, we are currently evaluating them and will present the results in another paper.

\subsubsection{Formal Property Verification}

FPV is the most commonly used app to verify the correct behaviour of the design. We have used FPV to verify several design features, such as FIFO and arbiters. Using a formal based approach helped expose several design bugs that would have been hard to find using the simulation based approach. The design also has several scan isolation based requirements where several signals need to be isolated to either logic 0 or logic 1 in the scan mode. This requirement was also verified with the FPV approach which gave us confidence in the design implementation.

The DUV also has an AHB bus that connects SPI to the registers. Due to time limitations, we did not prepare a dedicated UVC for the AHB since the correctness of it is implicitly verified with several UVM testcases. However, since AHB is an important aspect of the design and must comply with standards, we used Cadence ABVIP to verify its correctness. The AHB DUV acts as a subordinate and the ABVIP acts as a manager in the verification setup. Using this approach, we were able to verify the AHB bus.

\subsubsection{Clock Domain Crossing Verification}

The Clock Domain Crossing (CDC) app from Cadence Jasper offers formal based CDC and Reset Domain Crossing (RDC) verification. The tool is used to verify both structural as well as functional checks that are automatically generated based on several synchronizer schemes used in the design. Later, these checks/properties are also verified in the effect of metastability by using a Metastability Injection (MSI) flow. The real benefit of using the CDC app is that the user can also write custom checks/properties that can be proved in the metastability effects. Later, these properties and the MSI model can be exported to the simulation setup and reused while doing the verification. A detailed explanation of a pragmatic formal CDC verification is described in \cite{formal_cdc}.

\subsubsection{Unreachability Analysis}

The UNR app from Cadence Jasper is used to find out parts of the RTL code that are unreachable. This analysis helps to identify code coverage holes in an early cycle of RTL development. The analysis is shared with the design engineers to make sure either to fix the RTL code or waive them off if not needed. In the RADAR project, we were able to find a significant amount of unreachable code that was waived off before the tape-out.

\subsection{Verification Management}

The project uses a single point for the traceability of all requirements. All requirements and verification items (such as tests, checks and coverage) are added to the requirements management tool. Later, these verification items are mapped to the corresponding requirements. A script is used to export the vPlan in Cadence vManager readable format that is later used to map all the tests, checks, and coverage items from the regression runs. The requirements management tool acts as the one-stop shop for all traceability of requirements and their verification items. The regression setup in the project is divided on the basis of different strategies. We use a general regression setup for UVM simulations and others for CDC jitter, UPF, GLS and formal verification. All these regressions are combined for overall metrics analysis. A script was set to trigger automatic weekly regressions on the weekend and even nightly regressions basis at a point when we were very close to the tape-out.

To increase the verification throughput, we used an ML based regression approach that is offered by Cadence Xcelium ML tool. The tool generates an optimized Verification Session Input File (VSIF) based on the learning models derived out of the regression runs which contains a list of testcases that contribute the most to the coverage metrics. The results of the analysis to improve functional and code coverage with a similar or fewer number of regression runs are discussed in \cite{deepak_ml}. We also observed that the style of tests resulted in certain choices of tests in the optimized VSIF. The testcases are implemented to test either a particular functionality in isolation or a complex testcase that covers the isolation ones as well. In a usual regression, if the complex testcase fails, the isolation one would also fail. This helps to identify the actual root cause rather than debugging a more complex testcase. However, since the ML models are trained to pick up the best (and more complex) testcase that contributes more to the coverage, it naturally picks up the more complex testcases rather than the isolation ones. This does not limit the capabilities of ML, but rather something that is affected by the styling of tests. For future possibilities, we are also exploring the capabilities of Xcelium ML to expose bugs or unique failure signatures in less turnaround time.

\section{Results}
\label{results}
With the verification approach mentioned in this paper, we were successfully able to verify the chip within a short time frame. We achieved acceptable requirements coverage for the first stage, and the rest were waiveivable. Note that the extensive protocol checks were not part of the verification and are something planned for the next phase of the project. We also do not have exhaustive functional coverage for the first stage, which is planned for the next phase. Simulation based verification exhibited a high degree of coverage across various functional, performance, and stress tests. Connectivity verification confirmed the integrity of data paths and signals, while CDC verification mitigated potential metastability issues. The integration of ML based regression optimization contributed to achieving higher coverage in a shorter time frame. Using the approach we discussed, we achieved acceptable code coverage as well. Although the first stage of our verification environment had less comprehensive protocol checks and functional coverage, this level of code coverage compensates for those gaps. Overall, this approach strikes a good compromise between time and exhaustiveness.

\section{Conclusion}
\label{conclusion}
The paper presents a real case study of one of our recent projects based on RADAR sensors, and the goal was to efficiently verify the design in an aggressive timeline with high confidence. Our work highlights the importance of a multi-faceted verification strategy to tackle the challenges posed by complex SoC designs and presents the lessons learnt during the course of verification. A requirement-driven flow was also used to track coverage and backup verification holes with real coverage holes. The synergy between formal and simulation based methods, coupled with ML-driven optimization, provides a powerful framework for achieving comprehensive and efficient verification, ensuring the robustness and reliability of modern SoCs in the fast-paced consumer electronics landscape.

\printbibliography[heading=bibintoc]\label{sec:bibliography}%

\end{document}